\documentclass[PAPER, cernpreprint, atlasdraft=false, paper=letter, texlive=2023, UKenglish, texmf, orcidlogo]{atlasdoc}
 
\usepackage{atlaspackage}
\usepackage{atlasbiblatex}
\usepackage{multirow}
\usepackage{atlasphysics}
 
\ifthenelse{\boolean{AtlasDraft}}{
}{}
 
\graphicspath{{logos/}{figures/}}
 
\usepackage{ANA-TOPQ-2018-28-PAPER-defs}

\AtlasTitle{Measurement of the polarisation of $W$ bosons produced in top-quark decays using dilepton events at $\sqrt{s} = 13$~TeV with the ATLAS experiment}
 
\AtlasAbstract{
A measurement of the polarisation of $W$ bosons produced in top-quark decays is presented, using proton--proton collision
data at a centre-of-mass energy of $\sqrt{s} = 13$~TeV.
The data were collected by the ATLAS detector
at the Large Hadron Collider and correspond to an integrated luminosity of
139~fb$^{-1}$.
The measurement is performed selecting $t\bar{t}$ events decaying into final
states with two charged leptons (electrons or muons) and at least two $b$-tagged jets. The polarisation is extracted from
the differential cross-section distribution of the $\cos{\theta^{*}}$ variable, where
$\theta^{*}$ is the angle between the momentum direction of
the charged lepton from the $W$ boson decay and the
reversed momentum direction of the $b$-quark from the top-quark decay,
both calculated in the $W$ boson rest frame.
Parton-level results, corrected for the detector acceptance and resolution, are presented for the $\cos{\theta^{*}}$ angle.
The measured fractions of longitudinal, left- and right-handed polarisation
states are found to be $f_{0} = 0.684 \pm 0.005\,\mathrm{(stat.)}  \pm 0.014\,\mathrm{(syst.)}$, $f_{\mathrm{L}} = 0.318 \pm 0.003\,\mathrm{(stat.)} \pm 0.008\,\mathrm{(syst.)}$ and $f_{\mathrm{R}} = -0.002 \pm 0.002\,\mathrm{(stat.)} \pm 0.014\,\mathrm{(syst.)}$, in agreement with the Standard Model prediction.}
 
\AtlasRefCode{TOPQ-2018-28}
 
\PreprintIdNumber{CERN-EP-2022-197}

\AtlasJournalRef{Phys.\ Lett.\ B. 843 (2023) 137829}
\AtlasDOI{10.1016/j.physletb.2023.137829}

\hypersetup{pdftitle={ATLAS document},pdfauthor={The ATLAS Collaboration}}
 
\begin{document}
 
\maketitle

\section{Introduction}

\label{sec:introduction}
Discovered in 1995 by the CDF and D0 experiments~\cite{Abe:1995hr,D0:1995jca},
the top quark is the heaviest known elementary particle so far. Its abundant production
at the Large Hadron Collider (LHC) allows its
properties to be measured with unprecedented precision.
The properties of the top-quark decay vertex $Wtb$ are determined by the
$(V-A)$ structure of the weak interaction in the Standard Model (SM),
where $V$ and $A$ refer to the vector and axial-vector components of the
weak interaction, respectively.
The $Wtb$ vertex structure, and the masses of the three particles, govern
the decay properties of the $W$ boson produced in the top-quark decay.
In particular, they define the fractions of longitudinal (\fO), left-handed (\fL) and right-handed (\fR) polarised $W$ bosons, referred to as helicity fractions.
Calculations at next-to-next-to-leading order (NNLO) in quantum chromodynamics (QCD) yield the following values for the fractions: $\fO=0.687\pm0.005$, $\fL=0.311\pm0.005$ and $\fR=0.0017\pm0.0001$ assuming a top-quark mass
$\mtop = 172.8$~\gev, a $W$ boson mass $\mW = 80.401$~\GeV\
and a $b$-quark mass $\mb = 4.8$~\GeV~\cite{PhysRevD.81.111503}.
The uncertainties in the \fO and \fL fractions are dominated by the experimental uncertainties in the top-quark mass, while the uncertainty in \fR is dominated by uncertainties in the strong coupling constant and the $b$-quark mass.
This analysis tests the structure of the $Wtb$ vertex by measuring the helicity fractions of $W$ bosons produced in top-quark decays with high precision.
Precise measurements of these fractions can probe possible new physics processes~\cite{Etesami:2012bz} which modify the structure of the $Wtb$ vertex, such as dimension-six operators, introduced in effective field theories~\cite{Aguilar-Saavedra:2008nuh,PhysRevD.83.034006}.
The helicity fractions \fO and \fL are especially sensitive to the $C_{tW}$ Wilson coefficient~\cite{Brivio:2019ius}, and at the level of squared dimension-6 operators, $C_{bW}$ and $C_{\phi tb}$ also affect them.
Additionally, the expected value of \fR is very small, making it particularly sensitive to possible signs of new physics.
The $W\ell\nu$ vertex is assumed to follow the SM prediction, in accord with extensive studies of this vertex at LEP~\cite{Heister:730540, DELPHIWhel, L3:2004lwm, OPAL:2000mzv, LEP-2}.
 
The $W$ boson helicity fractions can be extracted from measurements of the
angular distribution of the decay products of the $W$ boson and the top quark.
The eight-component $W$ boson spin density matrix, with three spin-operator
and five tensor-operator components, entirely determines the angular
distribution of the products of the leptonic decay $W^{\pm} \to \ell^{\pm}\nu$, with
$\ell = e, \mu, \tau$. It can be expressed in terms of the polar and azimuthal angles
of the charged-lepton momentum in the $W$ boson rest frame. Integrating over the
azimuthal angle, the off-diagonal contributions vanish, and the normalised
differential distribution of the cosine of the polar angle $\thetas$, \ctheta,
depends on the helicity fractions as~\cite{WHelicity_Aguilar}
\begin{equation}
\label{equn:HelicityFractions}
\frac{1}{\sigma}\frac{\dif \sigma}{\dif \cos\theta^*} = \frac{3}{4}(1 - \cos^2\theta^*)f_0 +  \frac{3}{8}(1-\cos\theta^*)^2 f_\mathrm{L} + \frac{3}{8}(1+\cos \theta^*)^2 f_\mathrm{R}.
\end{equation}
The angle $\theta^*$ is defined as the angle between the momentum direction of
the charged lepton from the $W$ boson decay and the reversed
momentum direction of the $b$-quark from the decay of the top quark, both
calculated in the $W$ boson rest frame.
 
Previous measurements of the $W$ boson helicity fractions from the
ATLAS~\cite{TOPQ-2016-02}, CMS~\cite{CMS-TOP-13-008},
CDF~\cite{Aaltonen:2010ha}, and D0~\cite{Abazov:2010jn, Aaltonen:2012rz}
collaborations are in agreement with the SM predictions within their uncertainties.
The most recent ATLAS measurement was performed in the single-lepton channel of \ttbar
decays at 8~\tev~\cite{TOPQ-2016-02} and obtained
$\fO = 0.709 \pm 0.012$\,(stat.\,+\,bkg.)\,${}^{+0.015}_{-0.014}$\,(syst.),
$\fL = 0.299 \pm 0.008$\,(stat.\,+\,bkg.)\,${}^{+0.013}_{-0.012}$\,(syst.) and
$\fR = -0.008 \pm 0.006$\,(stat.\,+\,bkg.)\,$\pm\,0.012$\,(syst.) using the template fit method.
This result was combined with measurements from CMS
to yield the combined result
$\fO =  0.693 \pm 0.014$, $\fL = 0.315 \pm 0.011$ and $\fR = -0.008 \pm 0.007$, where
total uncertainties are quoted~\cite{TOPQ-2018-02}.
 
This Letter presents a measurement of the $W$ helicity fractions in the dileptonic final
state of the \ttbar pair.
Electron and muon final states are probed, including the leptonic decays of $\tau$-leptons.
This decay mode is chosen for its very small background contamination that is better understood than the backgrounds in the single-lepton channel, despite the more complicated kinematic reconstruction required in this final state.
The measurement uses the full set of ATLAS $pp$ collision data at $\sqs = 13~\tev$,
which corresponds to an integrated luminosity of $\SI{139}{\ifb}$.
The helicity fractions are extracted from a fit to the normalised differential
\ctheta distribution unfolded to parton level, in contrast to the previous ATLAS results quoted above, which measure the helicity fractions from a template fit to the detector-level distributions.
This provides a complementary measurement of the fractions while also measuring the differential cross-section distribution with respect to \ctheta.


\FloatBarrier
 
\section{The ATLAS detector}

\label{sec:detector}
ATLAS~\cite{PERF-2007-01,Abbott:2018ikt,CERN-LHCC-2012-009} is a multipurpose particle
detector designed with a forward--backward symmetric cylindrical geometry and nearly full \(4\pi\) coverage in solid angle.\footnote{\AtlasCoordFootnote}
It consists of an inner tracking detector (ID) surrounded by a thin superconducting solenoid providing a \SI{2}{\tesla} axial magnetic field, electromagnetic and hadronic calorimeters, and a muon spectrometer (MS).
The ID covers the pseudorapidity range \(|\eta| < 2.5\) and is composed of silicon pixel, silicon microstrip, and transition radiation tracking detectors.
Lead/liquid-argon (LAr) sampling calorimeters provide electromagnetic (EM) energy measurements with high granularity.
Hadronic calorimetry is provided by the steel/scintillator-tile calorimeter covering the central pseudorapidity range (\(|\eta| < 1.7\)).
The endcap and forward regions are instrumented with LAr calorimeters for both the EM and hadronic energy measurements up to \(|\eta| = 4.9\).
The MS surrounds the calorimeters and is based on three large air-core toroidal superconducting magnets with eight coils each.
The field integral of the toroids ranges between \num{2.0} and \SI{6.0}{\tesla\metre} across most of the detector.
The MS includes a system of precision tracking chambers and fast detectors for triggering.
A two-level trigger system is used to select events.
The first-level trigger is implemented in hardware and uses a subset of the detector information
to keep the accepted event rate below \SI{100}{\kHz}~\cite{TRIG-2016-01}.
This is followed by a software-based trigger that reduces the accepted event rate to \SI{1}{\kHz} on average.
An extensive software suite~\cite{ATL-SOFT-PUB-2021-001} is used in data simulation, in the reconstruction and analysis of real
and simulated data, in detector operations, and in the trigger and data acquisition systems of the experiment.


\FloatBarrier
 
\section{Data and simulated samples}

\label{sec:samples}
This measurement exploits proton--proton collision data recorded with the ATLAS detector from
2015 to 2018 at a centre-of-mass energy of $\sqrt{s} = 13~\tev$.
After applying data-quality requirements~\cite{Aad:2019hmz}, the data sample corresponds to an integrated
luminosity of 139\,fb$^{-1}$, determined by using the LUCID-2 detector~\cite{LUCID2} for the primary luminosity measurements.
Monte Carlo (MC) simulated samples are used in the analysis to optimise the event selection,
estimate the selection efficiency and predict contributions from various background processes.
 
The production of \ttbar events was modelled by
the \POWHEGBOX[v2]~\cite{Frixione:2007nw,Nason:2004rx,Frixione:2007vw,Alioli:2010xd} generator, using a
next-to-leading-order (NLO) matrix element (ME), a dynamic scale~\cite{ATL-PHYS-PUB-2015-011}, and an \hdamp parameter value of $1.5 \mtop$~\cite{ATL-PHYS-PUB-2016-020}.\footnote{The \hdamp parameter is a resummation damping factor and one of the parameters that controls the matching of \POWHEG matrix elements to the parton shower and thus effectively regulates the high-\pT radiation against which the \ttbar system recoils.}
The top-quark decay was modelled by a leading-order (LO) ME in \POWHEGBOX[v2],
with an approximate implementation of finite-width and interference effects in \PYTHIA[8.230]~\cite{Sjostrand:2014zea}.
The \ttbar sample is normalised to the cross-section prediction at NNLO in QCD, including the resummation of next-to-next-to-leading logarithmic (NNLL) soft-gluon terms calculated using
\TOPpp[2.0]~\cite{Beneke:2011mq,Cacciari:2011hy,Baernreuther:2012ws,Czakon:2012zr,Czakon:2012pz,Czakon:2013goa,Czakon:2011xx}.
This cross-section corresponds to \(\sigma(\ttbar)_{\text{NNLO+NNLL}} = 832 \pm 51\,\si{\pb}\) when using a top-quark mass of \(\mtop = \SI{172.5}{\GeV}\).
Production of a top quark in association with a $W$ boson ($tW$) was modelled by
\POWHEGBOX[v2] at NLO in QCD using the five-flavour scheme.
The diagram removal scheme~\cite{Frixione:2008yi} was used to remove interference and overlap with \ttbar production.
The inclusive cross-section is corrected to the theory prediction calculated at NLO in QCD with NNLL soft-gluon
corrections~\cite{Kidonakis:2010ux,Kidonakis:2013zqa}.
 
The production of $V$\;\!+\,jets ($V=Z,W$) events was simulated with the \SHERPA[2.2.1]~\cite{Bothmann:2019yzt} generator, using a NLO
ME for up to two partons, and a LO ME for up to four partons, calculated with
the Comix~\cite{Gleisberg:2008fv} and \OPENLOOPS~\cite{Buccioni:2019sur,Cascioli:2011va,Denner:2016kdg} libraries.
They were matched with the \SHERPA parton shower~\cite{Schumann:2007mg} using
the \MEPSatNLO prescription~\cite{Hoeche:2011fd,Hoeche:2012yf,Catani:2001cc,Hoeche:2009rj}.
The samples are normalised to a NNLO prediction~\cite{Anastasiou:2003ds}.
Samples of diboson final states ($VV$) were simulated with \SHERPA, including off-shell effects and Higgs boson contributions where appropriate.
Fully leptonic final states and semileptonic final states, where one boson decays leptonically and the other
hadronically, were generated using MEs at NLO accuracy in QCD for up to one additional parton and at LO accuracy
for up to three additional parton emissions.
The matrix-element calculations were matched and merged with the \SHERPA parton shower based on
Catani--Seymour dipole factorisation~\cite{Gleisberg:2008fv,Schumann:2007mg} using the \MEPSatNLO prescription.
The virtual QCD corrections were provided by the \OPENLOOPS library.
 
The production of $\ttbar V$ and $\ttbar H$ events was modelled at NLO
using the \MGNLO[2.3.3]~\cite{Alwall:2014hca} and \POWHEGBOX[v2] generators, respectively.
The simulated $\ttbar V$ and $\ttbar H$ events are normalised to the cross-sections
computed at NLO in QCD with the leading NLO electroweak
corrections~\cite{Campbell:2012dh,Frixione:2015zaa,deFlorian:2016spz}.
 
For all samples of simulated events, except those generated using \SHERPA, the decays of bottom and charm hadrons were performed by \EVTGEN~\cite{Lange:2001uf}.
In all processes, $\mtop$ was set to $172.5~\gev$. The \POWHEGBOX and \MGNLO
generators used the \NNPDF[3.0nlo] set of parton distribution functions (PDFs)~\cite{Ball:2014uwa}
and the events were interfaced to \PYTHIA~\cite{Sjostrand:2014zea}
to model the parton shower, hadronisation, and underlying event, with parameter values set according to the
A14 tune~\cite{ATL-PHYS-PUB-2014-021}. For events generated with \SHERPA, the \NNPDF[3.0nnlo] set of PDFs and a dedicated set of tuned parton-shower parameters were used.
After the event generation, the ATLAS detector response was simulated~\cite{SOFT-2010-01} using either a full detector simulation based on \GEANT~\cite{Agostinelli:2002hh}
or a faster parametric simulation~\cite{ATL-PHYS-PUB-2010-013} for the MC samples used to estimate the modelling uncertainties.
The effect of multiple interactions in the same or neighbouring bunch-crossings (pile-up) was
modelled by overlaying each hard-scattering event with simulated inelastic $pp$ events
generated by \PYTHIA[8.186]~\cite{Sjostrand:2007gs} using the \NNPDF[2.3lo] set of PDFs~\cite{Ball:2012cx} and parameter values set according to the A3 tune~\cite{ATL-PHYS-PUB-2016-017}.
 
The contribution from events with misreconstructed or non-prompt leptons passing the selection is estimated using MC samples for processes with
one prompt lepton in the matrix element. These include single-lepton \ttbar, $\ttbar V$, $\ttbar H$, $t$- and $s$-channel
single top quark, and $W$\;\!+\,jets production. After event selection, this background contribution is less than 1\% of the total predicted SM yield, including the \ttbar events.


\FloatBarrier
 
\section{Object and event selection}

\label{sec:selection}
 
Electron candidates are reconstructed from clusters of energy deposited in the electromagnetic calorimeter matched
to particle tracks inside the ID.
The candidates are identified with the \emph{TightLH} likelihood-based identification
criteria~\cite{ATL-PHYS-PUB-2017-022,EGAM-2018-01}.
They are required to have $\pt > 25~\gev$ and $|\eta| < 2.47$, excluding the transition region $1.37 < |\eta| < 1.52$ between the barrel
and endcaps.
Electron candidates must also have a transverse impact-parameter significance $|d_0/\sigma(d_0)| < 5$, measured relative to the beam line, and satisfy $|z_0 \sin{\theta}| < 0.5$~mm, where $\theta$ is the polar angle of the track and $z_0$ is the longitudinal distance from the primary vertex to the point where $d_0$ is measured.
Muon candidates are reconstructed from tracks in the MS matched to tracks in the ID.
The candidates are identified with the \emph{Medium} identification criteria~\cite{MUON-2018-03} with $\pt > 25~\gev$ and $|\eta| < 2.5$.
Additionally, muon candidates must satisfy $|z_0 \sin{\theta}| < 0.5$~mm  and $|d_0/\sigma(d_0)| < 3$.
Isolated electrons and muons are selected by requiring both the amount of energy deposited in close proximity in the calorimeters and the scalar sum of the transverse momenta of nearby tracks in the ID to be small.
 
Jet candidates are reconstructed from particle-flow objects~\cite{PERF-2015-09}, using the \antikt~\cite{Cacciari:2008gp} jet
algorithm with radius parameter $R=0.4$ implemented in the FastJet~\cite{Fastjet} software.
A jet energy scale calibration derived from 13~\tev\ data and simulation~\cite{JETM-2018-05} is applied to the reconstructed jets.
After the calibration, jet candidates are required to have $\pt > 25~\gev$ and $|\eta| < 2.5$.
To suppress jets originating from pile-up collisions, a `jet vertex
tagger' (JVT)~\cite{PERF-2014-03} discriminant requirement
is applied to jets with \pt below 60~\gev.
 
Jets containing $b$-hadrons are identified ($b$-tagged) using the DL1r algorithm~\cite{FTAG-2018-01,ATL-PHYS-PUB-2017-013}.
The algorithm combines inputs from the impact parameters of displaced vertices, as well as topological properties of secondary and tertiary vertices within a jet.
These inputs are then passed to a neural network that outputs three values, representing the probability of the jet being a light-flavour jet, a $c$-jet or a $b$-jet, which are then combined into a single discriminant.
The $b$-tagged jets are required to satisfy the operating point corresponding to 60\% efficiency for identifying $b$-quark jets in
simulated \ttbar\ events.
 
The missing transverse momentum vector $\vec{p}_{\textrm{T}}^{\text{\,miss}}$, with magnitude \met, is defined as the negative sum of the transverse momenta of the reconstructed and calibrated physical objects, as well as a `soft term' built from all other tracks that are associated with the primary vertex~\cite{PERF-2016-07}.
 
To avoid double-counting of detector signatures, overlapping physics objects are removed in the following order:
electrons sharing a track with a muon;
the closest jet within $\Delta R = 0.2$ of an electron;
electrons within $\Delta R = 0.4$ of a jet;
jets within $\Delta R = 0.4$ of a muon if they have at most two associated tracks;
muons within $\Delta R = 0.4$ of a jet.\footnote{For the overlap removal, $\Delta R$ is defined as $\Delta R \equiv \sqrt{\left(\Delta y\right)^2 + \left(\Delta \phi\right)^2}$,
where $y = (1/2)[(E+p_z)/(E-p_z)]$ is the rapidity of the object.}
 
Scale factors (SFs) are used to correct the efficiencies in simulation to those measured in data for
the electron and muon trigger, reconstruction,
identification, and isolation criteria~\cite{EGAM-2018-01, MUON-2018-03, TRIG-2018-05, TRIG-2018-01}.
Additionally, the energies of the electrons~\cite{EGAM-2018-01} and the \pt of the muons~\cite{PERF-2015-10} and jets~\cite{PERF-2016-04,ATL-PHYS-PUB-2015-015} are corrected using resonance decays.
SFs are also applied for the JVT requirement~\cite{ATLAS-CONF-2014-018} and for the $b$-tagging efficiencies for jets that
originate from the hadronisation of $b$-quarks~\cite{FTAG-2018-01}, $c$-quarks~\cite{FTAG-2020-08},
and $(u, d, s)$-quarks or gluons~\cite{ATLAS-CONF-2018-006}.
The amount of the pile-up in simulation is corrected to reproduce the number of reconstructed primary vertices in data.
 
\begin{figure}[!hbtp]
\centering
\subfloat[]{
\includegraphics[width=0.49\textwidth]{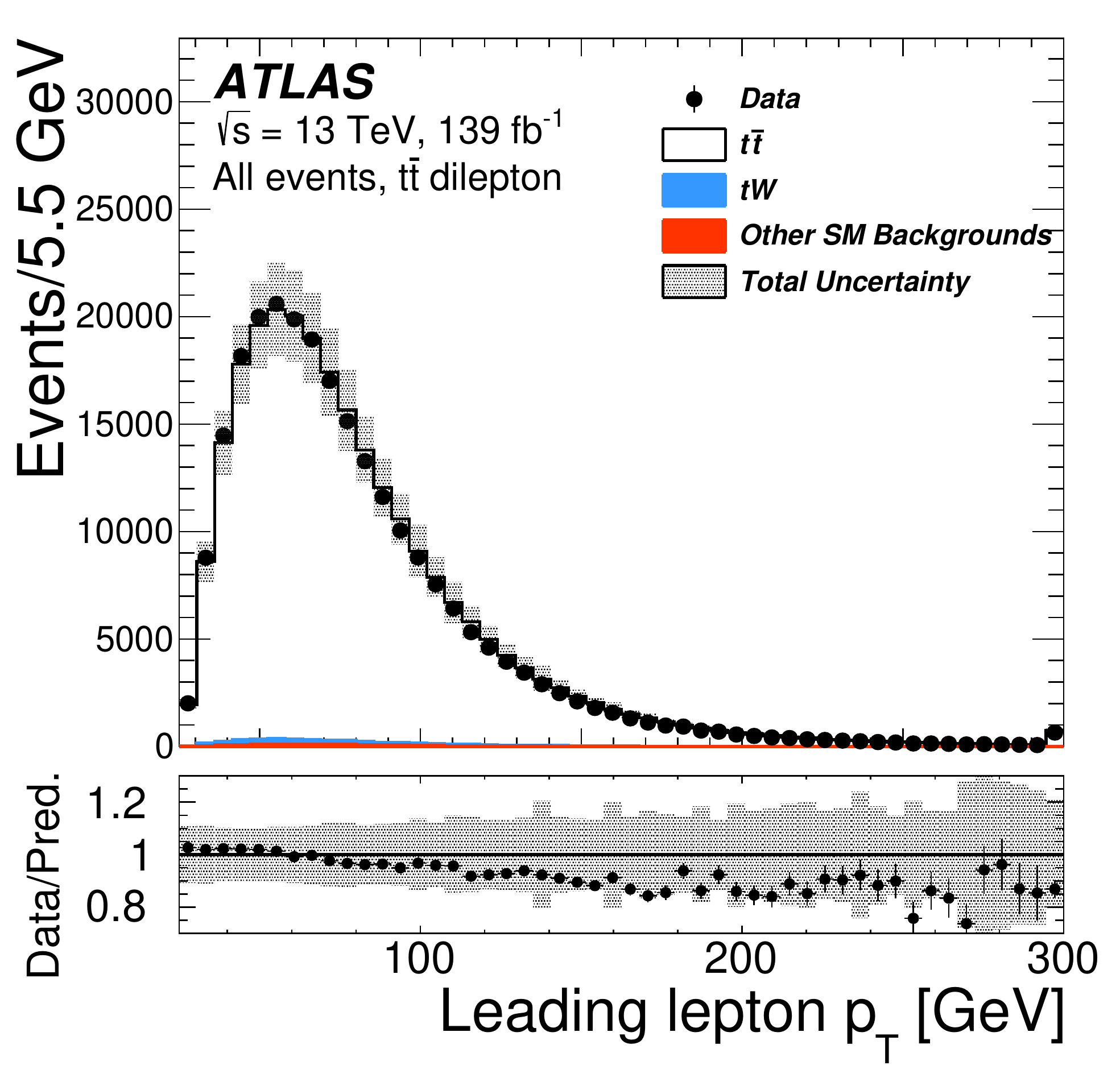}
\label{fig:control_lepton_pt}
}
\subfloat[]{
\includegraphics[width=0.49\textwidth]{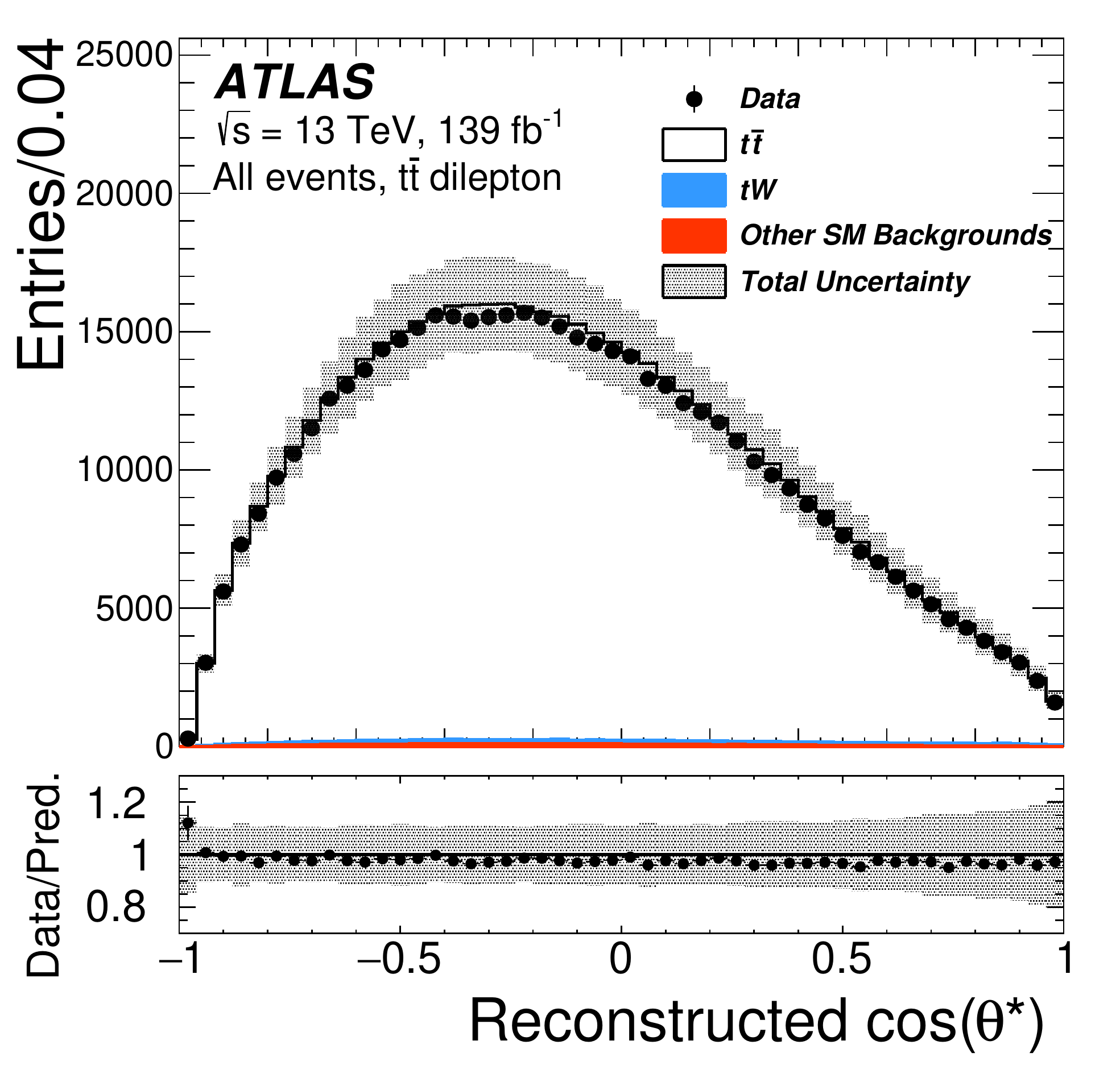}
\label{fig:control_cosT}
}
\caption{Comparison of observed data and predictions for (a) the \pT distribution of the leading lepton
and (b) the reconstructed \ctheta distribution containing measurements from both hemispheres of the \ttbar system.
All three lepton-flavour channels, \ee, \mumu and \emu, are combined. The hashed band represents the total uncertainty.
The bottom panel shows the ratio of data to prediction. The rightmost bins contain the overflow events. }
\label{fig:RecoDataMC}
\end{figure}

Events that satisfy the requirements of at least one of the single-electron or single-muon triggers~\cite{TRIG-2016-01,TRIG-2018-05,TRIG-2018-01} are selected.
The events are also required to have at least one reconstructed collision vertex with two or more associated tracks
with transverse momentum, \pt, greater than 500~\mev.
The vertex with the highest $\sum \pt^2$ of the associated tracks is taken as the primary vertex.
The selected events are required to have exactly two leptons (electrons or muons) of opposite electric charge
with $\pt > 25~\gev$ or $\pt > 27~\gev$ for the 2015 and 2016--2018 data-taking periods, respectively, to match
the increasing minimum \pt thresholds of the single-electron and single-muon triggers.
One of the reconstructed charged leptons must be matched to the lepton that passed the trigger requirement.
Additionally, the events are required to have at least two reconstructed jets with $\pt > 25~\gev$, with at least two of
these $b$-tagged at the 60\%-efficiency operating point. This `tight' operating point is chosen because it reduces the
background to a minimum level while keeping a large number of \ttbar events.
Furthermore, in same-flavour lepton events ($\ee$ and $\mumu$)
the invariant mass of the two charged leptons, $m_{\ell\ell}$, is required to be outside of the $Z$-boson mass window of 80--100~\gev,
and $\met > 60~\gev$ is required in order to suppress background originating from $Z$\,+\,jets events.
Finally, all events are required to have $m_{\ell\ell} > 15~\gev$ to suppress low-mass resonances.
After the event selection, the sample is expected to contain about 250\,000 \ttbar events. The background represents about 3.5\% of all events
passing the selection, with the largest contribution coming from the single-top $tW$ process.
Table~\ref{tab:EventYields_dil} shows the expected event yields after the event selection for the signal and background processes as well as the total number of events seen in data.
Figure~\ref{fig:RecoDataMC} compares the data with the predictions for the leading-lepton \pt\ distribution and
the \ctheta distribution after the event selection.
The observed deviation of the prediction from data in lepton \pt\ distribution arises from the top-quark \pt\ mismodelling by NLO generators~\cite{Beneke:2011mq,Cacciari:2011hy,Baernreuther:2012ws,Czakon:2012zr,Czakon:2012pz,Czakon:2013goa,TOPQ-2018-15} and is covered by the \ttbar modelling uncertainties.
 
\begin{table}[!h]
\centering
\caption{Event yields after the event selection for \ttbar, $tW$ and all other SM processes compared to data. The uncertainties on the predicted yields include all systematic uncertainties discussed in Section~\ref{sec:systematics}. \label{tab:eventYields}}
 
\label{tab:EventYields_dil}
\begin{tabular}{rccccc}
\toprule
Process & \ttbar & $tW$ & Other bkg. & Total prediction & Data \\
\midrule
Yields & 250000 $\pm$ 19000 & 4500 $\pm$ 800 & 2800 $\pm$ 1400 & 257000 $\pm$ 19000 & 251512\\
\bottomrule
\end{tabular}
\end{table}
 
\FloatBarrier
 
\section{Reconstruction of the top-quark kinematics}

\label{sec:reconstruction}
 
The observable sensitive to the \W boson helicity fractions, \ctheta, requires reconstruction of the kinematics of the \ttbar\ event
from the identified leptons, jets and missing transverse momentum.
The individual four-momenta of the two neutrinos from the \ttbar\ dileptonic decay are not measured in the detector but the sum of their transverse momenta, \met,
is measured and is used to reconstruct the top quark and top antiquark using the `neutrino weighting' (NW) method~\cite{Abbott:1998nw}. The method allows the
underconstrained system to be reconstructed by using the \W-boson and top-quark mass constraints and scanning over the neutrino pseudorapidities to find two possible
solutions for each of the assumed values of the neutrino pseudorapidities. The corresponding \met\ value is then compared with the measured \met\ of the event,
and an event weight based on the degree of their agreement, which takes into account the resolution of the measured \met, is computed.
For some events, no solution can be found, due to the finite resolution of the detector or mismeasurement of the input objects' transverse momenta. To mitigate this effect,
the transverse momenta of the measured jets are varied using a Gaussian function with a \pt-dependent width between 8\% and 14\% of the measured jet \pt~\cite{JETM-2018-05}.
This variation is repeated five times, increasing the probability of finding a solution. For the events with more than two $b$-tagged jets, the two $b$-jets with the highest \pt\
are used in the kinematic reconstruction. The top quark and top antiquark are reconstructed by assigning to their decays the $b$-jets and neutrino momenta corresponding to
the solution with the highest weight in the NW method. A solution is found for about 90\% of the \ttbar events.
Events where the NW method fails to find a solution are discarded and not used further.
The reconstruction efficiency for the background is lower, which helps to suppress it.
 
The statistical correlation of the distributions of \ctheta originating from the top quark and top antiquark was checked in the simulation and found to be small.
Thus, these two \ctheta distributions are combined into a single distribution used in the measurement.
Furthermore, since uncertainties related to lepton reconstruction are expected to be subdominant,
the \ctheta distributions from the three lepton-flavour channels, \ee, \mumu and \emu,
are also combined into a single distribution, mitigating statistical fluctuations.
Further investigation has shown that the increased sensitivity to systematic uncertainties, due to the different selection criteria imposed on the same-flavour channels, is not significant.


\FloatBarrier
 
\section{Differential cross-section}

\label{sec:differential}
In order to extract \W boson helicity fractions, the differential \ctheta cross-section is measured at parton level.
The parton level is defined as the full phase-space of the dileptonically decaying $W$ bosons from the \ttbar decay, including $\tau$-leptons.
The leptons are taken from the MC generator's `truth' record before final-state photon radiation, and in the case of $\tau$-leptons, also before decay.
The expected background, estimated using the MC simulation, is subtracted from the detector-level \ctheta distribution.
The background-subtracted detector-level \ctheta distribution is corrected for detector effects using an iterative Bayesian unfolding (IBU)
method~\cite{DAgostini:1994fjx} incorporated in the RooUnfold~\cite{Adye:2011gm} package with updated corrections to the
error propagation~\cite{Prosper:2011zz}.
At MC-truth level, both top quarks in a \ttbar\ event are required to decay leptonically.
 
The differential cross-section is calculated as
 
\begin{equation*}
\label{equn:unfolding}
\frac{\dif\sigma_{t\bar{t}}}{\dif\cos\theta^*_i} = \frac{1}{\mathcal{L}\cdot\Delta X_{i}\cdot \epsilon^{\text{sel}}_{i}} \cdot \sum_{j}R^{-1}_{ij}\cdot (N^{\text{obs}}_{j} - N_{j}^{\text{bkg}}),
\end{equation*}
 
\noindent where \textit{i} denotes a bin of the \ctheta distribution, $\Delta X_{i}$ is the width of bin $i$,
$\mathcal{L}$ is the integrated luminosity, and
$N_j^{\text{obs}}$ and $N_j^{\text{bkg}}$ are the observed number of data events and the estimated number of background events in bin \textit{j},
respectively. The $\epsilon^{\text{sel}}_{i}$ term corresponds to the probability for a MC-truth event to satisfy the reconstruction and selection criteria.
The migration matrix $R^{-1}_{ij}$ maps the binned parton-level events to
the binned detector-level events and is derived from simulated \ttbar events
decaying into dilepton final states, following the procedure
described in Ref.~\cite{TOPQ-2018-18}.
The probability of correct measurement of \ctheta is represented by the diagonal elements of the migration matrix, whereas the off-diagonal elements represent the probability of event migration between those bins.
 
The impact of the considered systematic uncertainties in the \ctheta distribution, described in Section~\ref{sec:systematics}, is estimated using pseudo-data obtained by systematically varying the detector-level distributions predicted by the simulation. Each varied distribution is then unfolded, and the difference between the MC-truth distribution and the unfolded distribution is considered
an uncertainty in the unfolded distribution. Different sources of systematic uncertainty are uncorrelated with each other, but each is correlated
across the bins.
 
\begin{figure}[!hbtp]
\centering
\subfloat[]{
\includegraphics[width=0.49\columnwidth]{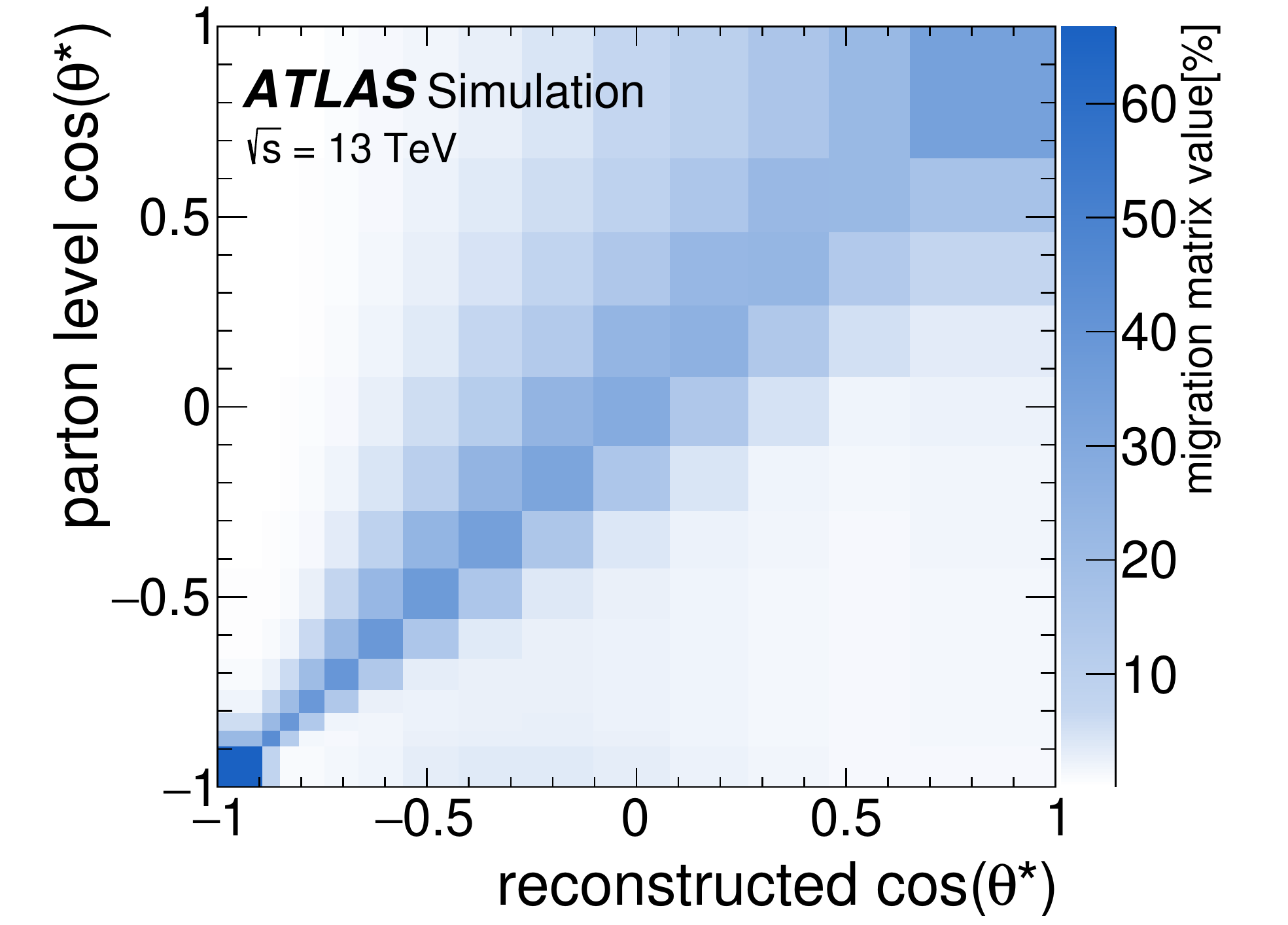}
\label{fig:migration_only}
}
\subfloat[]{
\includegraphics[width=0.49\columnwidth]{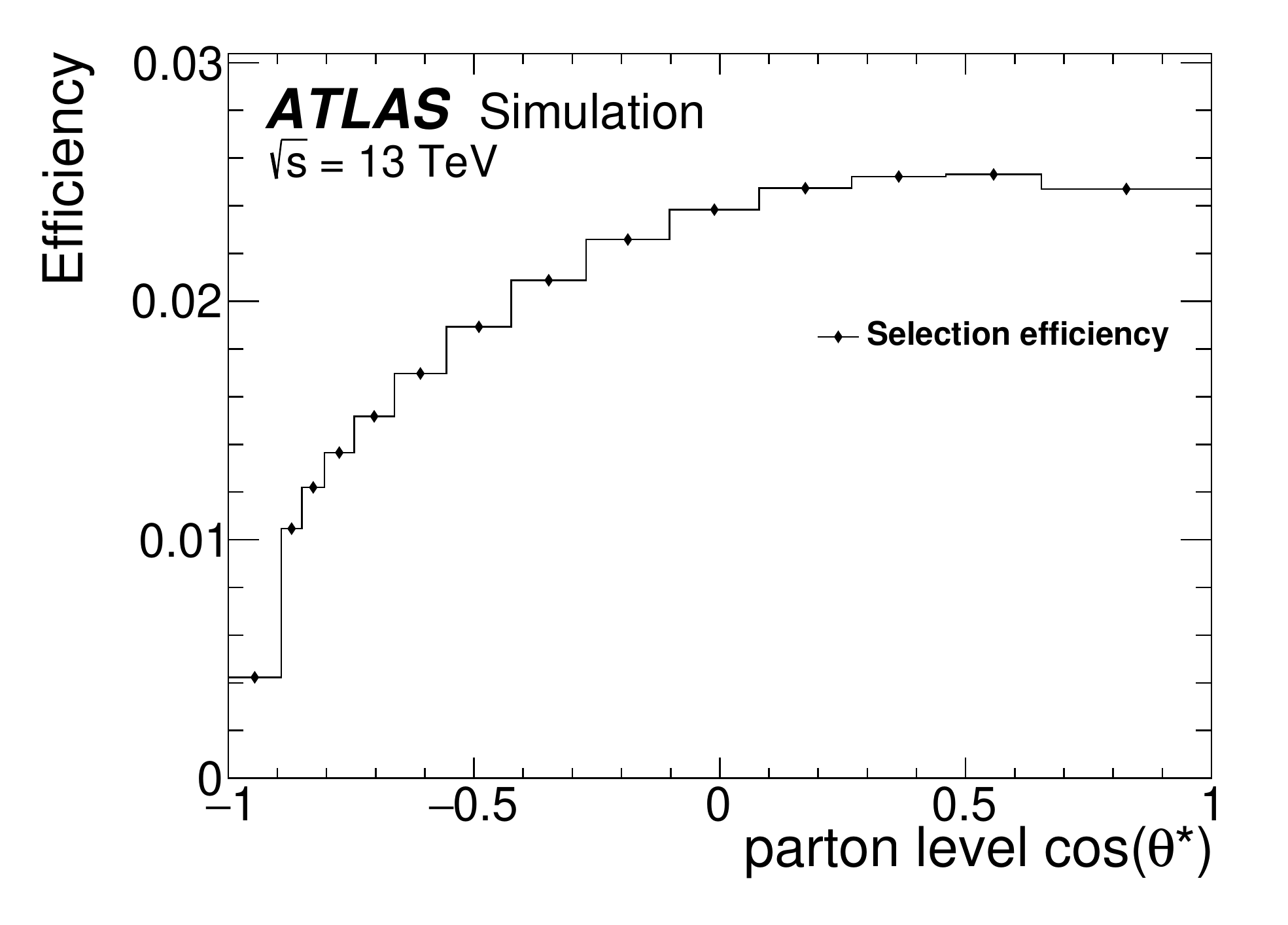}
\label{fig:selEff}
}
\caption{Migration matrix (Figure~\ref{fig:migration_only}) and selection efficiency per bin (Figure~\ref{fig:selEff}). The migration matrix is obtained as a result of the bin optimisation.
The entries represent probabilities (expressed as percentages) for an event with $\ctheta$ in bin $i$ at parton level to have reconstructed $\ctheta$ in bin $j$ at
detector level. The selection efficiency is calculated with respect to the true \ttbar dilepton events in the bin. The error bars on the selection efficiency points are too small to be seen in the distribution.\label{fig:migration}}
\end{figure}

The binning of the \ctheta distribution is optimised to mitigate statistical fluctuations. It is chosen such that
each bin of the detector-level \ctheta distribution contains at least 1.5\% of the total number of events
and, furthermore, at least 30\% of each bin's events originate from the corresponding MC-truth bin.
This procedure results in the migration matrix and selection efficiency shown in Figure~\ref{fig:migration}.
 
MC simulated events are used to validate the unfolding method and the extraction of the $W$ boson helicity fractions.
However, it was found that the MC-truth \ctheta distribution deviates slightly from the
quadratic formula in Eq.~(\ref{equn:HelicityFractions}) after the simulation of the parton shower.
The formula is followed exactly at the matrix-element level, but the four-momentum-reshuffling in the parton shower generator
distorts the distribution at a few per mille level.
To circumvent this problem and construct MC samples with well-defined true helicity fractions,
the MC-truth \ctheta distribution is reweighted to match the functional form of Eq.~(\ref{equn:HelicityFractions})
when using the values of the helicity fractions calculated at NNLO in QCD~\cite{PhysRevD.81.111503}. The weights are derived
from a ratio of the MC-truth \ctheta distribution with a thousand bins and the analytic function in
Eq.~(\ref{equn:HelicityFractions}) with fractions set to the NNLO prediction.
 
Stress tests to validate the unfolding method are performed using simulated \ctheta distributions representing pure helicity states.
These distributions are obtained for each fraction, using the reweighting procedure described in Ref.~\cite{TOPQ-2016-02} to obtain
\ctheta distributions corresponding to different values of the helicity fractions.
Several stressed distributions are constructed with values of helicity fractions in ranges of \fL and \fR
corresponding to $2\sigma$ variations of the measured fractions in Ref.~\cite{TOPQ-2016-02}.
The stressed distributions are unfolded and compared with a similarly stressed parton-level
distribution to check the linearity and for potential biases.
The latter are minimised by optimising the IBU regularisation parameter that controls the number of iterations in the unfolding algorithm.
The optimal value is found to be 180, for which the observed bias is of the order of the expected statistical uncertainty of the measurement.
This rather large regularisation parameter is due to the combination of migration matrix having large off-diagonal elements and a non-linear \ctheta distribution.


\FloatBarrier
 
\section{Extraction of the helicity fractions}

\label{sec:fractions}
The helicity fractions are extracted by fitting Eq.~(\ref{equn:HelicityFractions}) to the measured normalised differential cross-section distribution
and minimising the \chisq defined as
 
\begin{equation}
\label{equn:chi2constrained}
\chi^2 = \Delta y^\text{T} C^{-1} \Delta y,
\end{equation}
 
\noindent where $\Delta y$ is a vector containing the differences between the bin yield in data and the
value of the function in Eq.~(\ref{equn:HelicityFractions}).
Due to the unitarity constraint on the sum of the helicity fractions, the parameter representing one of the fractions is replaced with one minus the sum of the other helicity fractions.
 
For each bin, the function is evaluated
at the point along the horizontal axis where the expected bin yield matches the value of the function
defined by Eq.~(\ref{equn:HelicityFractions}). This choice is motivated by the non-linear shape
of the analytic function.
The matrix $C$ is the covariance matrix of the normalised differential cross-section distribution and contains both
the statistical and systematic uncertainties. The covariance matrix is generated according to the procedure described in Ref.~\cite{TOPQ-2018-18}.
Statistical and systematic uncertainties are estimated using `toy' experiments on the pre-unfolded detector-level \ctheta distribution in data and then propagating the variations to the unfolded distribution.
The statistical uncertainty is estimated by applying independent Poisson fluctuations to the individual bins.
Systematic uncertainties are estimated using Gaussian smearing of individual bins by the corresponding systematic variation.
All of these changes to the distribution are summed in quadrature to estimate the total uncertainty per bin.
\par
Since a normalised distribution is used in the fit, one degree of freedom is removed by
removing one bin of the \ctheta distribution from Eq.~(\ref{equn:chi2constrained}).
The minimisation is carried out using the MINUIT program~\cite{minuit}.
The parameters representing the helicity fractions are not restricted in the fit, and are allowed to take unphysical values outside of the $[0,1]$ interval.
 
Linearity tests of the helicity fraction extraction procedure
are performed with stressed distributions generated as described in Section~\ref{sec:differential}.
For each stressed distribution, the \chisq defined by Eq.~(\ref{equn:chi2constrained}) is minimised
with the covariance matrix computed when including only the MC statistical uncertainty, and the extracted values of
the helicity fractions are compared with their input values.
Any non-closure seen is included as a systematic uncertainty of the measurement.


\FloatBarrier
 
\section{Systematic uncertainties}

\label{sec:systematics}
 
Systematic uncertainties may affect the selection efficiency for the \ttbar signal, the bin migrations, the number of events expected from the background processes, and the shapes of the background distributions.
These effects are estimated by varying each source of systematic uncertainty by one standard deviation and considering the resulting deviation from the nominal expectation as the uncertainty.
For the extraction of the helicity fractions, the systematic uncertainties enter the covariance matrix that is used in the fit as described in Section~\ref{sec:fractions}.
 
The effects of uncertainties in the modelling of the \ttbar signal are estimated by independently varying the renormalisation and
factorisation scales in the matrix element by factors of 0.5 and 2, but normalising the signal to the nominal cross-section.
A variation of the Var3c parameter of the A14 tune~\cite{ATL-PHYS-PUB-2017-007}, which impacts the renormalisation scale for initial-state radiation, is considered independently.
An uncertainty from the final-state radiation modelling is estimated by doubling and halving the nominal renormalisation scale for emissions from the parton shower.
The uncertainty due to the choice of the \hdamp\ parameter value is estimated by increasing its value by a factor of two and symmetrising the impact.
All the \ttbar\ scale variations described above and the \hdamp\ variation are referred to as ``\ttbar\ radiation" in Table~\ref{tab:groupedFit}.
Additionally, an uncertainty due to the choice of parton shower generator is estimated by comparing the nominal MC sample with an alternative MC sample that uses  \HERWIG[7.1.3]~\cite{Bahr:2008pv,Bellm:2015jjp} with the H7UE set of tuned parameters~\cite{Bellm:2015jjp} and the \MMHT[lo] PDF set~\cite{Harland-Lang:2014zoa} instead of the \PYTHIA[8.230] generator.
Furthermore, an uncertainty due to the nominal choice of generator is estimated by comparing the nominal prediction with the prediction from \MGNLO[2.6.0]~\cite{Alwall:2014hca} interfaced with \HERWIG[7.1.3].
The uncertainty due to using a top-quark mass of $172.5~\gev$ is estimated by raising and lowering the mass by 0.5~\gev, which is approximately the uncertainty in the measurement of the top-quark mass by the ATLAS Collaboration~\cite{TOPQ-2013-02}.
The PDF uncertainty is estimated by considering the internal variations of the PDF4LHC~\cite{Butterworth:2015oua} PDF set.
For all \ttbar\ modelling uncertainties, all predictions are reweighted to match the NNLO helicity fractions in the full phase-space as described in Section~\ref{sec:differential}.
The impact of the limited size of the simulated samples on the extraction of the helicity fractions is negligible.
 
For the $tW$ process, effects of uncertainties in the renormalisation and factorisation scales, the Var3c parameter value in the A14 tune, and final-state radiation modelling are estimated following the same procedure as used for the \ttbar signal.
Additionally, an uncertainty due to the nominal choice of parton shower and hadronisation generator is estimated by comparing the nominal MC prediction with a prediction using \HERWIG[7.0.4] instead of the \PYTHIA[8.230] generator.
An uncertainty due to the choice of generator is estimated by comparing distributions from \MGNLO[2.6.0] with the nominal ones from \POWHEGBOX[v2].
Furthermore, an uncertainty due to the overlap between the $tW$ and \ttbar\ processes is estimated by comparing samples using the
diagram removal scheme with those using the diagram subtraction scheme~\cite{Frixione:2008yi}.
 
An uncertainty of 5.3\%, estimated from the scale and PDF variations~\cite{Kidonakis:2010ux}, is applied to the $tW$ background normalisation.
A conservative 50\% cross-section uncertainty is used for $\ttbar V$, $\ttbar H$, $Z$\,+\,jets, and $VV$ production and for processes with non-prompt leptons.
This conservative uncertainty was found to have negligible impact on the final result.
 
An uncertainty of 1.7\% in the integrated luminosity is considered for all processes~\cite{ATLAS-CONF-2019-021}.
The uncertainty due to pile-up is determined by varying the average number of interactions per bunch-crossing by 3\% in the simulation.
Uncertainties in the calibration, reconstruction and identification of the different reconstructed objects are also considered.
For electrons and muons, these include the uncertainties in the measured SFs for triggering, reconstruction, identification and isolation~\cite{EGAM-2018-01, MUON-2018-03, TRIG-2018-05, TRIG-2018-01}, as well as in the electron- and muon-momentum calibration and resolution~\cite{EGAM-2018-01,PERF-2015-10}.
For hadronic jets, the uncertainties in the jet energy scale (JES)~\cite{PERF-2016-04} and jet energy resolution (JER)~\cite{ATL-PHYS-PUB-2015-015}, as well as the uncertainties in the SFs for the JVT~\cite{ATLAS-CONF-2014-018} and the tagging of jets as $b$-jets~\cite{FTAG-2018-01,FTAG-2020-08,ATLAS-CONF-2018-006}, are considered.
All uncertainties associated with reconstructed objects are propagated to the \met and an uncertainty in the soft term is also considered~\cite{PERF-2016-07}.
The JES and JER uncertainties are determined using a model with 30 and 8 independent components, respectively.
The uncertainties in the $b$-tagging calibration include nine/five/five independent variations for the $b$-/$c$-/light-jet calibrations
and two components for the MC-based uncertainty extrapolation to very high \pt jets.


\FloatBarrier
 
\section{Results}

\label{sec:results}
 
\begin{figure}[!htb]
\centering
\subfloat[]{
\includegraphics[width=0.49\textwidth]{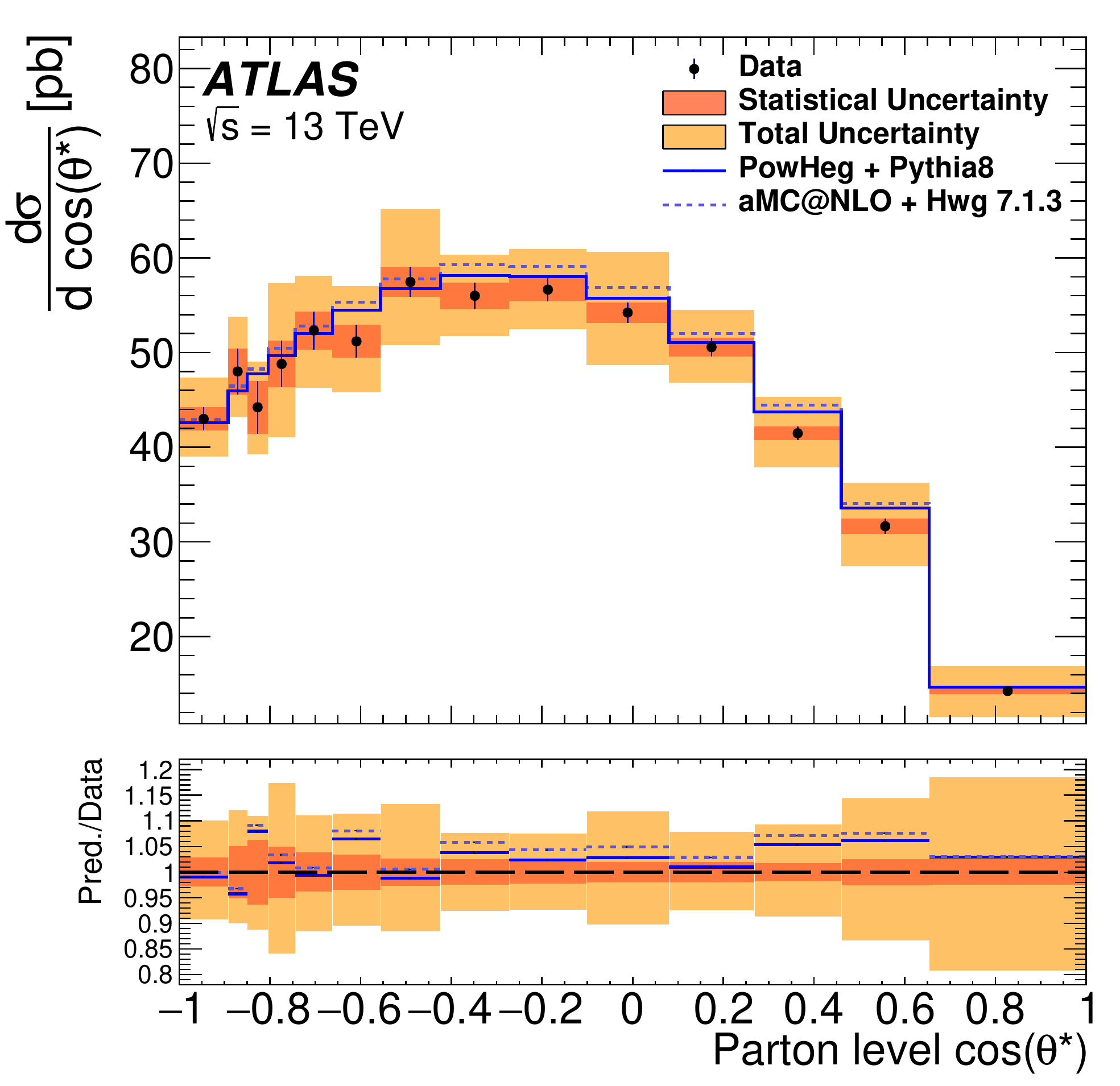}
\label{fig:diffXSec_single}
}
\subfloat[]{
\includegraphics[width=0.49\textwidth]{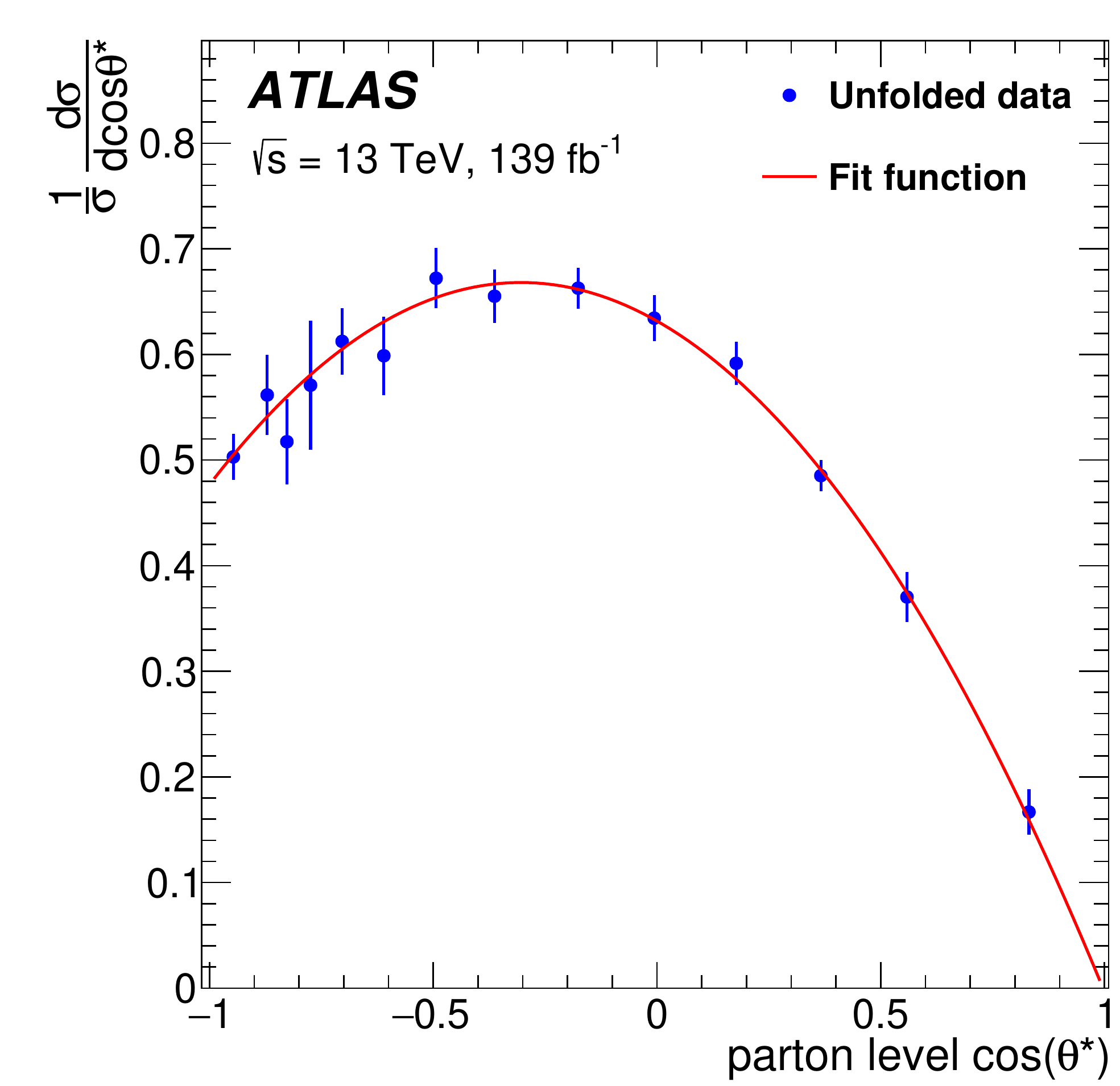}
\label{fig:fitResults}
}
\caption{The unfolded \ctheta differential distribution (Figure~\ref{fig:diffXSec_single}) and the unfolded normalised \ctheta distribution (Figure~\ref{fig:fitResults}).
The total statistical and systematic uncertainty per bin is shown in Figure~\ref{fig:diffXSec_single}.
The parton-level distribution predicted by \POWHEGBOX interfaced with \PYTHIA is shown. An uncertainty originating from the limited number of simulated events is
included in the prediction but is not visible.
The unfolded normalised \ctheta distribution in data is shown with the function of Eq.~(\ref{equn:HelicityFractions})
overlaid, using the helicity fractions \fO, \fL and \fR determined from the fit. The total uncertainties are shown on data points.
}
 
\label{fig:fit_fractions}
\end{figure}

Figure~\ref{fig:diffXSec_single} shows unfolded \ctheta differential distribution with statistical and total uncertainties
compared with the prediction of \ttbar MC simulation. The unfolded
normalised \ctheta distribution along with the fit function used to measure the helicity fractions \fO, \fL and \fR is shown in Figure~\ref{fig:fitResults}.
In the fit, \fL and \fR are free parameters and the \fO parameter is set to $\fO = 1 - \fL - \fR$ to preserve the unitarity
of the sum of the three parameters.
The helicity fractions are found to be
\begin{equation*}
\begin{split}
\fO &= \phantom{-}0.684 \pm 0.005~\text{(stat.)}\pm 0.014~\text{(syst.)}, \\
\fL &= \phantom{-}0.318 \pm 0.003~\text{(stat.)} \pm 0.008~\text{(syst.)}, \\
\fR &= -0.002 \pm 0.002~\text{(stat.)} \pm 0.014~\text{(syst.)}, \\
\end{split}
\end{equation*}
with the covariance and correlation matrices of the fit shown in Table~\ref{tab:covarPostFit}.
The difference in the magnitude of the statistical uncertainty on the three helicity fractions is caused by the negative statistical correlations between the fractions.
The $\chi^{2}$ divided by the number of degrees of freedom of the fit is 0.267, demonstrating good agreement between the fitted functional form and data corrected to parton level.
The covariance matrix is estimated from a $2 \times 2$ matrix obtained directly from the fit, with a third row and column calculated analytically from the unitarity
constraint on the helicity fractions.
The obtained values and uncertainties, including the covariance matrix, do not change if any two of the parameters are chosen as free parameters and the third one
is calculated analytically.
 
The expected uncertainties in the helicity fraction obtained from a fit to the MC predictions are identical to the measured uncertainties.
 
As a further test of the unfolding procedure, a \ctheta distribution is generated in the Monte Carlo, based on the measured values of the helicity fractions; this distribution is unfolded, helicity fractions are extracted, and compared to the input values.
A small degree of non-closure is observed but the corresponding uncertainty is negligible and does not change the total uncertainties listed in Table~\ref{tab:groupedFit}.
 
\renewcommand{\arraystretch}{1.5}
\begin{table}[!htb]
\centering
\caption{Covariance matrix and correlation matrix for the measured helicity fractions.}
\begin{tabular}{rccc|ccc}
\toprule
\empty & \multicolumn{3}{c}{Covariance} & \multicolumn{3}{c}{Correlation} \\
\midrule
& \fO & \fL & \fR & \fO & \fL & \fR \\
\midrule
\fO & $\phantom{-}2.125 \times 10^{-4}$ & $-3.665 \times 10^{-5}$ & $-1.758 \times 10^{-4}$ & $1$ & $-0.308$ & $-0.841$ \\
\hline
\fL & $-3.665 \times 10^{-5}$ & $\phantom{-}6.651 \times 10^{-5}$ & $-2.986 \times 10^{-5}$ & $-0.308$ & $1$ & $-0.255$ \\
\hline
\fR & $-1.758 \times 10^{-4}$ & $-2.986 \times 10^{-5}$ & $\phantom{-}2.057 \times 10^{-4}$ & $-0.841$ & $-0.255$ & $1$ \\
\bottomrule
\label{tab:covarPostFit}
\end{tabular}
 
\end{table}
\renewcommand{\arraystretch}{1.0}
The impact of different categories of systematic uncertainty and the data's statistical uncertainty on the \fO, \fL and \fR measurement
is summarised in Table~\ref{tab:groupedFit}. They are estimated by generating a covariance matrix which includes all sources
of uncertainty except for the considered category and repeating the fit.
The considered category contributes an uncertainty whose square is the difference of the squares of nominal-fit and repeated-fit symmetrised total uncertainties for each helicity fraction.
The systematic uncertainty dominates the total uncertainty for all three helicity fractions.
The largest systematic uncertainty in all three helicity fractions arises from the modelling of \ttbar\ production, and
is dominated by the uncertainty from the choice of matrix-element generator.
Other significant uncertainties come from the jet energy scale and resolution as well as electron and muon reconstruction.
\begin{table}[!htb]
\centering
\caption{Impact of different categories of systematic uncertainty and the data's statistical uncertainty on the \fO, \fL and \fR measurement.
The squares of the quoted numbers are evaluated as the difference of the squares of the nominal-fit total uncertainties (quoted in the last row)
and those extracted from a fit using a covariance matrix including all sources of uncertainty except for the considered category.
In the fits with partial covariance matrices, the best-fit values for the helicity fractions differ between the fits with the full covariance matrix by up to a quarter of the total uncertainty.
The shifts in the best-fit values impact the precision of the uncertainty estimate for the categories.
The total uncertainty is different from the sum in quadrature of the different components
because of correlations between different uncertainties entering the covariance matrix.}
\begin{tabular}{rccc}
\toprule
\textbf{Category} & $\sigma_{\fO}$ & $\sigma_{\fL}$ & $\sigma_{\fR}$ \\
\midrule
\multicolumn{4}{c}{\textbf{Detector modelling}} \\
\midrule
Jet reconstruction & 0.008 & 0.004 & 0.010 \\
Flavour tagging & 0.003 & 0.001 & 0.001 \\
Electron reconstruction & 0.003 & 0.002 & 0.002 \\
Muon reconstruction & 0.003 & 0.003 & $<10^{-3}$\\
\met (soft term) & $<10^{-3}$ & 0.002 & $<10^{-3}$ \\
Pile-up & 0.002 & 0.002 &  $<10^{-3}$ \\
Luminosity & 0.001 & 0.001 & $<10^{-3}$ \\
\midrule
\multicolumn{4}{c}{\textbf{Signal and background modelling}} \\
\midrule
$t\bar{t}$ generator choice & 0.009 & 0.004 & 0.004 \\
$t\bar{t}$ parton shower/hadronisation & 0.003 & 0.001 & 0.002 \\
$t\bar{t}$ radiation & 0.007 & 0.002 & 0.007 \\
Top-quark mass & 0.005 & 0.003 & 0.007 \\
PDF & 0.002 & 0.001 &  $<10^{-3}$ \\
Single top production &  $<10^{-3}$ & 0.002 & $<10^{-3}$ \\
Other background  & 0.002 & 0.001 & $<10^{-3}$ \\
\midrule
Total systematic uncertainty & 0.014 & 0.008 & 0.014 \\
Data statistical uncertainty & 0.005 & 0.003 & 0.002 \\
\midrule
Total uncertainty & 0.015 & 0.008 & 0.014 \\
\bottomrule
\end{tabular}
\label{tab:groupedFit}
\end{table}
 
The impact on the extracted helicity fractions arising from the mis-modeling of the top-quark \pt\ distribution in the MC simulation was tested by correcting the top-quark \pt\ in simulation to match the calculation at NNLO in QCD with NLO electroweak corrections~\cite{Czakon:2017wor}.
No significant effect on the expected uncertainty was observed.


\FloatBarrier
 
\section{Conclusion}

\label{sec:conclusion}
A measurement of the $W$ boson helicity fractions using \ttbar events in the dilepton
final state is presented. It used data from 13~\tev\ $pp$
collisions collected by the ATLAS detector at the LHC, corresponding to an integrated luminosity of 139~fb$^{-1}$.
The fractions are extracted from the normalised
differential cross-section distribution of \ctheta corrected to the parton level.
This provides a complementary measurement of the helicity fractions to the previously published ATLAS results.
The measured fractions of longitudinal, left- and right-handed polarisation states are found to be $\fO = \fzero$, $\fL = \fleft$ and $\fR = \fright$, in agreement within one standard deviation with the Standard Model calculation at NNLO in QCD.


\FloatBarrier
 
\section*{Acknowledgements}


We thank CERN for the very successful operation of the LHC, as well as the
support staff from our institutions without whom ATLAS could not be
operated efficiently.
 
We acknowledge the support of
ANPCyT, Argentina;
YerPhI, Armenia;
ARC, Australia;
BMWFW and FWF, Austria;
ANAS, Azerbaijan;
CNPq and FAPESP, Brazil;
NSERC, NRC and CFI, Canada;
CERN;
ANID, Chile;
CAS, MOST and NSFC, China;
Minciencias, Colombia;
MEYS CR, Czech Republic;
DNRF and DNSRC, Denmark;
IN2P3-CNRS and CEA-DRF/IRFU, France;
SRNSFG, Georgia;
BMBF, HGF and MPG, Germany;
GSRI, Greece;
RGC and Hong Kong SAR, China;
ISF and Benoziyo Center, Israel;
INFN, Italy;
MEXT and JSPS, Japan;
CNRST, Morocco;
NWO, Netherlands;
RCN, Norway;
MEiN, Poland;
FCT, Portugal;
MNE/IFA, Romania;
MESTD, Serbia;
MSSR, Slovakia;
ARRS and MIZ\v{S}, Slovenia;
DSI/NRF, South Africa;
MICINN, Spain;
SRC and Wallenberg Foundation, Sweden;
SERI, SNSF and Cantons of Bern and Geneva, Switzerland;
MOST, Taiwan;
TENMAK, T\"urkiye;
STFC, United Kingdom;
DOE and NSF, United States of America.
In addition, individual groups and members have received support from
BCKDF, CANARIE, Compute Canada and CRC, Canada;
PRIMUS 21/SCI/017 and UNCE SCI/013, Czech Republic;
COST, ERC, ERDF, Horizon 2020 and Marie Sk{\l}odowska-Curie Actions, European Union;
Investissements d'Avenir Labex, Investissements d'Avenir Idex and ANR, France;
DFG and AvH Foundation, Germany;
Herakleitos, Thales and Aristeia programmes co-financed by EU-ESF and the Greek NSRF, Greece;
BSF-NSF and MINERVA, Israel;
Norwegian Financial Mechanism 2014-2021, Norway;
NCN and NAWA, Poland;
La Caixa Banking Foundation, CERCA Programme Generalitat de Catalunya and PROMETEO and GenT Programmes Generalitat Valenciana, Spain;
G\"{o}ran Gustafssons Stiftelse, Sweden;
The Royal Society and Leverhulme Trust, United Kingdom.
 
The crucial computing support from all WLCG partners is acknowledged gratefully, in particular from CERN, the ATLAS Tier-1 facilities at TRIUMF (Canada), NDGF (Denmark, Norway, Sweden), CC-IN2P3 (France), KIT/GridKA (Germany), INFN-CNAF (Italy), NL-T1 (Netherlands), PIC (Spain), ASGC (Taiwan), RAL (UK) and BNL (USA), the Tier-2 facilities worldwide and large non-WLCG resource providers. Major contributors of computing resources are listed in Ref.~\cite{ATL-SOFT-PUB-2021-003}.


\clearpage
 
\printbibliography
 
\clearpage
 
\begin{flushleft}
\hypersetup{urlcolor=black}
{\Large The ATLAS Collaboration}

\bigskip

\AtlasOrcid[0000-0002-6665-4934]{G.~Aad}$^\textrm{\scriptsize 102}$,
\AtlasOrcid[0000-0002-5888-2734]{B.~Abbott}$^\textrm{\scriptsize 120}$,
\AtlasOrcid[0000-0002-7248-3203]{D.C.~Abbott}$^\textrm{\scriptsize 103}$,
\AtlasOrcid[0000-0002-1002-1652]{K.~Abeling}$^\textrm{\scriptsize 55}$,
\AtlasOrcid[0000-0002-8496-9294]{S.H.~Abidi}$^\textrm{\scriptsize 29}$,
\AtlasOrcid[0000-0002-9987-2292]{A.~Aboulhorma}$^\textrm{\scriptsize 35e}$,
\AtlasOrcid[0000-0001-5329-6640]{H.~Abramowicz}$^\textrm{\scriptsize 151}$,
\AtlasOrcid[0000-0002-1599-2896]{H.~Abreu}$^\textrm{\scriptsize 150}$,
\AtlasOrcid[0000-0003-0403-3697]{Y.~Abulaiti}$^\textrm{\scriptsize 117}$,
\AtlasOrcid[0000-0003-0762-7204]{A.C.~Abusleme~Hoffman}$^\textrm{\scriptsize 137a}$
% ...
\bigskip
\\

% ...
$^{29}$Physics Department, Brookhaven National Laboratory, Upton NY; United States of America.\\
% ...
$^{35}$$^{(a)}$Facult\'e des Sciences Ain Chock, R\'eseau Universitaire de Physique des Hautes Energies - Universit\'e Hassan II, Casablanca;$^{(b)}$Facult\'{e} des Sciences, Universit\'{e} Ibn-Tofail, K\'{e}nitra;$^{(c)}$Facult\'e des Sciences Semlalia, Universit\'e Cadi Ayyad, LPHEA-Marrakech;$^{(d)}$LPMR, Facult\'e des Sciences, Universit\'e Mohamed Premier, Oujda;$^{(e)}$Facult\'e des sciences, Universit\'e Mohammed V, Rabat;$^{(f)}$Institute of Applied Physics, Mohammed VI Polytechnic University, Ben Guerir; Morocco.\\
% ...
$^{55}$II. Physikalisches Institut, Georg-August-Universit\"{a}t G\"ottingen, G\"ottingen; Germany.\\
% ...
$^{102}$CPPM, Aix-Marseille Universit\'e, CNRS/IN2P3, Marseille; France.\\
$^{103}$Department of Physics, University of Massachusetts, Amherst MA; United States of America.\\
% ...
$^{117}$Department of Physics, New York University, New York NY; United States of America.\\
% ...
$^{120}$Homer L. Dodge Department of Physics and Astronomy, University of Oklahoma, Norman OK; United States of America.\\
% ...
$^{137}$$^{(a)}$Departamento de F\'isica, Pontificia Universidad Cat\'olica de Chile, Santiago;$^{(b)}$Millennium Institute for Subatomic physics at high energy frontier (SAPHIR), Santiago;$^{(c)}$Instituto de Investigaci\'on Multidisciplinario en Ciencia y Tecnolog\'ia, y Departamento de F\'isica, Universidad de La Serena;$^{(d)}$Universidad Andres Bello, Department of Physics, Santiago;$^{(e)}$Instituto de Alta Investigaci\'on, Universidad de Tarapac\'a, Arica;$^{(f)}$Departamento de F\'isica, Universidad T\'ecnica Federico Santa Mar\'ia, Valpara\'iso; Chile.\\
% ...
$^{150}$Department of Physics, Technion, Israel Institute of Technology, Haifa; Israel.\\
$^{151}$Raymond and Beverly Sackler School of Physics and Astronomy, Tel Aviv University, Tel Aviv; Israel.\\
% ...
$^{a}$ Also Affiliated with an institute covered by a cooperation agreement with CERN.\\
$^{b}$ Also at An-Najah National University, Nablus; Palestine.\\
$^{c}$ Also at Borough of Manhattan Community College, City University of New York, New York NY; United States of America.\\
$^{d}$ Also at Bruno Kessler Foundation, Trento; Italy.\\
$^{e}$ Also at Center for High Energy Physics, Peking University; China.\\
$^{f}$ Also at Center for Interdisciplinary Research and Innovation (CIRI-AUTH), Thessaloniki ; Greece.\\
% ...

\end{flushleft}


\end{document}